\documentclass{aiaa-tc}
\RequirePackage[latin1]{inputenc}
\usepackage{url}
\usepackage{amsmath}
\usepackage{amssymb}
\usepackage{cite}
\usepackage{psfrag}
\usepackage{epsfig}
\usepackage{epstopdf}
\usepackage{pstricks}
\usepackage{subfigure}
\usepackage{lettrine}
\usepackage{verbatim}
\usepackage{array}
\usepackage{varwidth}
\usepackage{multicol}
\usepackage{booktabs}
\usepackage{latexsym}
\usepackage{eucal}
\usepackage{upref}

\newcommand{\wt}[1]{{\widetilde{#1}}}

\def\d{\textup d}

\def\CO2{\mathrm{CO_2}}

\def\o2{\mathrm{O_2}}
\def\n2{\mathrm{N_2}}
\def\H2{\mathrm{H_2}}
\def\n{\mathrm{N}}
\def\o{\mathrm{O}}

\def\h2o{\mathrm{H_2O}}
\def\h2{\mathrm{H_2}}
\def\oh{\mathrm{OH}}
\def\ch4{\mathrm{CH_4}}

\newcommand\textsubscript[1]{\ensuremath{{}_{\text{#1}}}}

\newenvironment{list-}
{\begin{list}{--\ }{\setlength{\rightmargin}{\leftmargin}}}
{\end{list}}

\newcommand\frds{\displaystyle\frac}

\newcommand{\ppder}[2]{\frds{\partial #1}{\partial #2}}

\def\d{\textup d}
\newcommand\zvar{{Z^{''^2}}}

\title{Numerical investigation of high-pressure  combustion in rocket engines
using Flamelet/Progress-variable models}

\author{
A. Coclite\thanks{Ph.D Student,  a.coclite@poliba.it} \\
{\normalsize\itshape Politecnico di Bari, Dipartimento di Meccanica, Matematica e Management, DIMMM}\\
{\normalsize\itshape Centro di Eccellenza in Meccanica Computazionale, CEMeC}\\
L. Cutrone\thanks{Research engineer, Propulsion Unit, AIAA Senior Member,
l.cutrone@cira.it, +390823623108}\\
{\normalsize\itshape Centro Italiano Ricerche Aerospaziali, CIRA}\\ 
\and
P. De Palma\thanks{Professor, AIAA member, depalma@poliba.it, +390805963226 },
G. Pascazio\thanks{Professor, pascazio@poliba.it, +390805963221}\\
{\normalsize\itshape Politecnico di Bari, Dipartimento di Meccanica, Matematica e Management, DIMMM}\\
{\normalsize\itshape Centro di Eccellenza in Meccanica Computazionale, CEMeC}\\
}

\AIAApapernumber{2015-????}
\AIAAconference{45th AIAA/ASME/SAE/ASEE Joint Propulsion Conference \& Exhibit}
\AIAAcopyright{\AIAAcopyrightD{2015}}

\makeglossary

\begin{document}
\maketitle

\section*{Abstract}
The present paper deals with the numerical study of high pressure 
H2/L$\o2$ combustion for propulsion systems. The present research effort is
driven by the continued interest in achieving low cost, reliable
access to space and more recently, by the renewed interest in hypersonic 
transportation systems capable of reducing time-to-destination.
Moreover, combustion at high pressure has been assumed as a 
key issue to achieve better propulsive performance and lower environmental 
impact, as long as the replacement of hydrogen with a hydrocarbon, to 
reduce the costs related to ground operations (propellant handling,
infrastructure and procedures) and increase flexibility.

Starting from this background, the current work provides a model for 
the numerical simulation of high-pressure turbulent combustion employing 
detailed chemistry description, embedded in a Reynolds averaged 
Navier-Stokes equations solver with a Low Reynolds number $k-\omega$ 
turbulence model.
The model used to study such a  combustion phenomenon is an extension of the standard flamelet--progress-variable
(FPV) turbulent combustion model combined with a Reynolds Averaged Navier-Stokes equation Solver
(RANS). In the FPV model, all of the thermo-chemical quantities are evaluated by evolving the mixture fraction $Z$ and a progress variable $C$.
When using a turbulence model in conjunction with FPV model, a probability density function (PDF) is 
required to evaluate statistical averages (e.g., Favre average) of chemical quantities. 
The choice of such PDF must be a compromise between computational costs and accuracy level. 
State-of-the-art FPV models are built presuming the functional shape of the joint PDF of $Z$ and $C$ in 
order to evaluate Favre-averages of thermodynamic quantities.
The model here proposed evaluates the most probable joint distribution of $Z$ and $C$ 
without any assumption on their behavior with the Statistically Most Likely Distribution (SMLD) framework. 
This provides a more general model in the context of FPV approach.

\section{Introduction}

The development of new technologies to enhance and control combustion processes is nowadays fundamental in order to increase efficiency and reduce emissions in many engineering applications, such as reciprocating internal combustion engines; advanced gas turbine systems, 
pulse detonation engines, high-speed air-breathing propulsion devices. 
Hydrogen is one of the preferred fuel because of its properties in terms of very short ignition delay time and high energy per unit weight. 
The investigation of hydrogen combustion presents significant difficulties and high costs either following the experimental approach or the numerical one.
Moreover, high-Reynolds-number turbulent combustion is a formidable multi-scale problem, where the interaction between chemical kinetics, molecular, and turbulent transport occurs over a wide range of length and time scales. These features pose severe difficulties in the analysis and comprehension of the basic phenomena involved in supersonic combustion.\\ 
In this context, the simulation of turbulent reacting flows is very useful to cut down experimental costs and to advance the comprehension of the basic physical mechanisms. 
Turbulent combustion is a multi-scale problem, where the interaction between chemical kinetics, molecular, and turbulent transport occurs over a wide range of length and time scales. 
The numerical simulation of such phenomena with detailed chemistry is today still prohibitive. The huge computational cost stems from the fact that, even for a simple fuel, detailed kinetic mechanism can involve thousands of reactions and, consequently, hundreds of chemical species. 
In recent years, the need for efficient tools has driven the research towards: i) developing models for turbulent combustion in order to understand and accurately mimic the interaction between turbulence and chemistry~\cite{heinz2010,fan2012,jin2013dns,Cecere2012}; 
ii) studying improved kinetic schemes to describe the combustion process~\cite{boivin2012supersonic,saxena2006,williams2008,Bezgin2013}.
Moreover, simplified approaches to combustion modeling have been proposed to further reduce the number of equations to be solved; for instance, the reduction of the chemical scheme in intrinsic low dimensional manifolds~(ILDM)~\cite{maas}; the flamelet-based approaches such as the flamelet--progress-variable~(FPV)~\cite{pierce} or flame prolongation of ILDM~(FPI)~\cite{laminarhydrogen}; and Flamelet Generated Manifolds approach~(FGM)~\cite{oijen}. 

An additional modeling difficulty is related to real-gas effects in high-pressure conditions.
For high-pressure propulsion systems, the pressure of the injected fluid is always supercritical, whereas its temperature may be either subcritical or supercritical. However, the fluid is typically injected into a chamber in which both the pressure and the temperature exceed the critical values for both the fuel and the oxidizer, so that a fast transition to the supercritical state is observed~\cite{mayer98}, and the liquid phase can be described as a dense gaseous jet. At such operating conditions, the ideal-gas equation of state cannot predict the correct p-v-t relation for the oxidizer and the fuel; for example, the density of oxygen predicted using the ideal-gas equation of state at supercritical conditions can be one fourth of the real value. Moreover, real-gas effects have a significant impact on the flame structure in high-pressure combustion, as shown in Ref~(\citen{gamal2000}) for the case of a premixed hydrogen/oxygen flame, and in Ref~(\citen{ribert08}) for that of counterflow hydrogen/oxygen diffusion flame. Therefore, a suitable equation of state, together with adequate constitutive equations for the transport properties, may be warranted.

The present work is based on the FPV model for non-premixed turbulent flames with a particular choice for the probability density function (PDF) needed to evaluate the thermo-chemical Favre averages combined with an an Eulerian single-phase numerical method for computing the mixing and combustion of liquid propellants at operating conditions typical of rocket combustion chambers which computes
the detailed chemistry for cryogenic applications efficiently and introduces a real-gas model in the flamelet library, so as to compute high-pressure combustion accurately~\cite{CutroneCAF09}.
The aim of this work is to study the applicability of the statistically most likely distribution (SMLD)~\cite{pope} approach to model joint-PDF of $Z$ and $\Lambda$ in the case of high-pressure combustion. The proposed joint-SMLD approach is very interesting since it represents a good compromise between computational costs and accuracy level.
The results obtained using the proposed model are validated versus experimental data and compared with numerical results obtained using the standard FPV model.


\section{The flamelet--progress-variable models}
For the case of non-premixed combustion of interest here, the basic assumptions of the flamelet model are fulfilled
for sufficiently large Damk\"{o}hler number, $D_{a}$. In fact, when the reaction zone thickness is very thin with respect to the Kolmogorov
length scale, turbulent structures are unable to penetrate into the reaction zone and cannot destroy the laminar flame structure. 
Effects of turbulence only result in a deformation and
straining of the flame sheet and locally the flame structure can be described as function of the mixture fraction,  $Z$,
the scalar dissipation rate, $\chi$, and the time. The scalar dissipation rate is a measure of the gradient of the mixture fraction 
representing the molecular fluxes of the species towards the flame and is defined as 
$\chi=2D_Z(\nabla Z)^2$, where $D_Z$ is the molecular diffusion coefficient of the chemical species.
Therefore, the entire flame behavior can be obtained as a combination of solutions of the laminar flamelet equation.
In the present work we consider a further simplification assuming a steady flamelet behavior, so that chemical effects are entirely
determined by the value of $Z$, whereas $\chi$ describes the effects of the flow on the flame structure according to
the following steady laminar flamelet equation (SLFE) for the generic variable $\phi$:
\begin{equation}
\label{slfe}
{-\rho\frac{\chi}{2}\frac{\partial^2 \phi}{\partial Z^2}=\dot\omega_\phi}.
\end{equation}
In equation \eqref{slfe},  $\rho$ is the density and  $\dot\omega_\phi$ is the source term related to $\phi$~\cite{piercemoin2004}, different from zero in the case of finite rate chemistry. 
In particular, in this work, the FPV model proposed by Pierce and Moin~\cite{pierce,piercemoin2004} is employed to evaluate all of the thermo-chemical quantities involved in the combustion process. This approach is based on the parameterization of the generic thermo-chemical quantity, $\phi$, in terms of the mixture fraction, $Z$, and of the progress parameter, $\Lambda$, instead of $\chi$:
\begin{equation}
\label{phi}
{\phi=F_\phi(Z,\Lambda)}.
\end{equation}
In principle a transport equation for $\Lambda$ can be solved, however, it has several unclosed terms that need to be modeled~\cite{IhmeCF1}. So that, it is defined a progress variable, $C$, obtained trough: $C=F_C(Z,\Lambda)$. A transport equation for $C$ is solved and the flamelet library is so parameterized in terms of $Z$ and $C$ assuming that the $F_c$, see equation~\eqref{phi}, is invertible, 
\begin{equation}
\label{lambda_C}
{\Lambda=F^{-1}_C(Z,C)}\, .
\end{equation}   
Equation~\eqref{phi}, in terms of $Z$ and $C$, reads:
\begin{equation}	
\label{phi_lambda}
{\phi=F_\phi(Z,F^{-1}_C(Z,C))}\, .
\end{equation}
The choice of the progress variable is not unique and some recent works discuss in details this issue proposing a procedure for its optimal 
selection~\cite{ihmejcp2012,cuenotCF2012,vervischCF2013}. A suitable definition for the progress variable 
is the sum of the mass fraction of the main products~\cite{ihmejcp2012}; for hydrogen combustion:
\begin{equation}
{C=Y_{H_2O}}.
\end{equation} 
Equation~\eqref{phi} is taken as the solution of the SLFE~\eqref{slfe}. 
The solution variety over $\chi = \chi_{st}$ is  called S-curve.
A key difficulty to integrate the flamelet equation is to know a priori informations about the scalar dissipation rate dependence on the mixture fraction, $\chi(Z)$. 
In fact, flamelet libraries are computed in advance and are assumed to be independent of the flow field.
Therefore, the dependence of $\chi$ on $Z$ has to be modeled~\cite{pitsch98,pitschchenpeters98,kimwilliams93}. 
In this work, the functional form of $\chi(Z)$ has been taken from an idealized flow configuration, as proposed by
Peters~\cite{peters84}; the distribution of the scalar dissipation rate in a counterflow diffusion flame, $\Phi(Z)$,  is employed, scaled in the following way:
\begin{equation}
\label{chi_st}
{\chi(Z)=\chi_{st}\frac{\Phi(Z)}{\Phi(Z_{st})}},
\end{equation}
where $\chi_{st}$ and $Z_{st}$ are evaluated at the stoichiometric point~\cite{peters84}.
From equation \eqref{phi} one can obtain the Favre-averaged value of $\phi$ and of its variance using the definitions:
\begin{equation}
\label{media}
{\widetilde\phi=\int\int F_\phi(Z,\Lambda)\widetilde{P}(Z,\Lambda)dZ d\Lambda},
\end{equation}
\begin{equation}
\label{varianza}
{\widetilde{\phi''^2}=\int\int (F_\phi(Z,\Lambda)-\widetilde\phi)^2\widetilde{P}(Z,\Lambda)dZd\Lambda},
\end{equation}
where $\widetilde P(Z,\Lambda)$ is the density-weighted PDF,
\begin{equation}
{\widetilde P(Z,\Lambda)=\frac{\rho P(Z,\Lambda)}{\overline{\rho}}},
\end{equation}
$P(Z,\Lambda)$ is the joint PDF and $\overline{\rho}$ is the Reynolds-averaged density. As usual, $\phi$ can be decomposed as:
\begin{equation}
\phi=\widetilde \phi+ \phi''\, ,
\qquad
\widetilde \phi = \frac{\overline{\rho \phi }}{\overline{\rho}}\, , 
\qquad
\text{and} 
\qquad
\rho=\overline{\rho} +\rho'\, ,
\end{equation}
where $\phi''$ and $\rho'$ are the fluctuations.
This function plays a crucial role in the definition of the model, affecting both its accuracy and computational costs.
Moreover, the choice of such a function is not straightforward because of the unknown statistical behavior of the two variables $Z$ and $\Lambda$~\cite{IhmeCF1}.
The definition of $\widetilde P(Z,\Lambda)$ is still an open problem whose solution is being pursued by several researches~\cite{ihmeal2005,IhmeCF2,DeMeesterCF2012,AbrahamPoF2012}.
The aim of this work is to provide a more general model based on the statistically most likely distribution (SMLD)~\cite{pope} for the joint PDF of $Z$ and 
$\Lambda$~\cite{cocliteAIMETA2013}; this model evaluates the statistical correlation between such variables and, using the total energy conservation to evaluate the temperature field, represents a effective method for the simulation of compressible reacting flows. 
Here, the performance of such a combustion model are assessed by employing
a Reynolds-Averaged Navier--Stokes solver with $k$-$\omega$ turbulence model~\cite{CutroneCAF09} to compute a hydrogen-air supersonic combustion.

\subsection{Presumed probability density function model}
\label{par:pdf}
In this section, the standard FPV model~\cite{piercemoin2004} (called here Model~A) and the joint-PDF SMLD model (called here Model~B)~\cite{cocliteAIMETA2013} are briefly described. 

Since the mixture fraction, $Z$, and the progress parameter, $\Lambda$, are the two independent variables of the combustion model, they form a basis by which one can derive all of the thermo-chemical quantities. 
When a turbulence model is used, the evaluation of the average quantities, see equations \eqref{media} and \eqref{varianza}, requires the PDF
to be known or somehow presumed.
Such a PDF establishes the statistical correlation between $Z$ and $\Lambda$. 
Employing the Bayes' theorem, 
\begin{equation}
\label{bayes}
{\widetilde{P}(Z,\Lambda)=\widetilde P(Z)\widetilde P(\Lambda|Z)} \, ,
\end{equation}
one usually presumes the functional shape of the marginal PDF of $Z$ and of the conditional PDF of $\Lambda|Z$.
In model~A, the basic assumption is the statistical independence between $Z$ and $\Lambda$, so that $\widetilde{P}(Z,\Lambda)=\widetilde P(Z)\widetilde P(\Lambda)$.
Furthermore, the statistical behavior of the mixture fraction is described by a $\beta$~-~distribution.
In fact, even though the definition of $\widetilde P(Z)$ is still an open question~\cite{pope}, 
it has been shown by several authors that the mixture fraction behaves like a passive scalar whose statistical distribution can be approximated by a $\beta$~function~\cite{cook,jimenez,wall}. 
The two parameter family of the $\beta$-distribution in the interval $x\in [0,1]$ is given by:
\begin{equation}
\label{beta}
{\beta(x;\widetilde{x},\widetilde{x''^2})=x^{a-1}(1-x)^{b-1}\frac{\Gamma(a+b)}{\Gamma(a)\Gamma(b)}},
\end{equation}
where $\Gamma(x)$ is the Euler function and $a$ and $b$ are two parameters related to $\widetilde x$ and $\widetilde{x''^2}$
\begin{equation}
\label{aeb}
{a=\frac{\widetilde x(\widetilde x- \widetilde{x}^2-\widetilde{x''^2})}{\widetilde{x''^2}}, \ \ \ b=\frac{(1-\widetilde x)(\widetilde x-\widetilde{x}^2-\widetilde{x''^2})}{\widetilde{x''^2}}}.
\end{equation} 
Moreover, $\widetilde P(\Lambda)$ is chosen as a Dirac distribution, implying a great simplification in the theoretical framework.
With these criteria, the Favre-average of a generic thermo-chemical quantity is given by:
\begin{equation}
\label{phidelta}
{\widetilde\phi=\int\int F_\phi(Z,\Lambda)\widetilde \beta(Z)\delta(\Lambda-\widetilde{\Lambda})dZd\Lambda=\int F_\phi(Z,\widetilde \Lambda)\widetilde \beta(Z)dZ}.
\end{equation} 
Therefore, one has to solve only three additional transport equations (for $\widetilde Z$, $\widetilde{Z''^2}$ and $\widetilde \Lambda$) to evaluate all of the thermo-chemical quantities, thus avoiding the expensive solution of one transport equation for each chemical species. 

Model~B, based on the SMLD approach to model the joint PDF, does not need any assumption about the form of $\widetilde{P}(Z,\Lambda)$.
Following such an approach, the probability distribution can be evaluated as a function of an arbitrary number of moments of $Z$ and $\Lambda$.
It is noteworthy that, even though equation~\eqref{slfe} is based on the assumption that $Z$ and $\Lambda$ are independent, one can properly take into account the statistical correlation between $Z$ and $\Lambda$ employing the SMLD joint-PDF in the evaluation of the effects of turbulence~\cite{ihmeal2005}. 

In this work, the first three moments of the joint probability density function $\widetilde P(\vec x)$, where $\vec x=(Z,\Lambda)^T$, are assumed to be known; 
therefore, the joint-PDF reads~\cite{cocliteAIMETA2013}:
\begin{multline}
\label{smld2}
\widetilde P_{SML,2}(Z,\Lambda)= \frac{1}{\mu_0}\exp\Bigl\{-\Bigl[\mu_{1,1} (Z-\widetilde Z)+\mu_{1,2}(\Lambda-\widetilde \Lambda)\Bigr]\\-\frac{1}{2}\Bigl[\mu_{2,11}(Z-\widetilde Z)^2+\mu_{2,12}(Z-\widetilde Z)(\Lambda-\widetilde \Lambda)+\mu_{2,21}(\Lambda-\widetilde \Lambda)(Z-\widetilde Z)+\mu_{2,22}(\Lambda-\widetilde \Lambda)^2 \Bigl]
\Bigr\}.
\end{multline}
In the equation above, $\mu_0$ is a scalar, $\vec{\mu_1}$ is a two~-~component vector,
and $\overleftrightarrow{\mu_2}$ is a square matrix of rank two:
\begin{eqnarray}
\label{moltiplicatori}
\mu_0&=&\int d\vec x \widetilde P_{SML,2}(\vec x) ,\\
\label{moltiplicatori1}
-\mu_{1,i}&=&\int d\vec x\partial_{x_i} \widetilde P_{SML,2}(\vec x)=\beta(1;\widetilde\xi_i,\widetilde{\xi_i''^2})-\beta(0;\widetilde \xi_i,\widetilde{\xi_i''^2}) ,\\
\label{moltiplicatori2}
\delta_{kl}-\mu_{2,kn}\ \widetilde{\xi'_n\xi'_l}&=&\int d\vec x \partial_{x_k}((x_l-\widetilde\xi_l)\widetilde P_{SML,2}(\vec x))=\beta(1;\widetilde \xi_k,\widetilde{\xi'_k\xi'_l})-\widetilde\xi_k\mu_{1,l} ,
\end{eqnarray} 
where $i$, $k$, $n$, and $l$ indicate the vector components; 
$\widetilde\xi_i$, $\widetilde\xi'_i$ and $\widetilde{\xi^{''2}_i}$ are the mean ($\widetilde{x_i}$), the fluctuation ($x_i-\widetilde{x_i}$) and the variance ($\widetilde{x_i''^2}$) of the $i$-th component of $\vec x$, respectively; finally, $\beta$ indicates the beta distribution function.

In Model~B,
one has to solve four additional transport equations (for $\widetilde Z$, $\widetilde{Z''^2}$, $\widetilde C$, and $\widetilde{C''^2}$) 
to evaluate all of the thermo-chemical quantities.

\section{Governing equations}
\subsection{Thermodynamics}
\label{par:thermo-model} 
The Peng--Robinson (PR) Equation of State (EoS)~\cite{pr76,reidbook} is employed in the present work: 
\begin{equation}
p=\frac{RT}{V_m-b}-\dfrac{a}{(V_m^2+2V_mb-b^2)},
\label{eq:ceospr}
\end{equation} 
where $R$ is the universal gas constant and $V_m$ is the molar volume.
The parameters $a(T)$ and $b$ account for the effects of 
attractive and repulsive forces between molecules,  a proper 
temperature dependence of $a$ being essential for predicting vapor pressures correctly~\cite{twu91}.
The PR EoS is one of the most frequently used cubic equations of state,
due to its straightforward implementation and its accuracy~\cite{harstad97}.
The extension to a multi-component mixture, according to the mixing rules proposed by
Reid \emph{et al.}~\cite{reidbook}, provides the following expressions
of the parameters $a$ and $b$:
\begin{equation}
a=\sum_i^{N_s}\sum_j^{N_s}\chi_i\chi_j a_{ij}(T),\qquad b=\sum_i^{N_s}\chi_i b_i,
\end{equation} %
where $N_s$ is the number of species,
$\chi_i$ is the molar fraction of the species \emph{i}, and the coefficients
$a_{ij}(T)$ and $b_{i}$ are given in Refs~(\citen{harstad97,cutroneCTR}) and are omitted here for brevity.

To ensure the self-consistency of the model, all of the thermodynamic 
properties of the flow are calculated from the 
same equation of state. The properties of interest
for the present fluid dynamic simulations are the specific 
enthalpy, $h$, and the constant-pressure specific heat, $c_p$.
Such properties can be obtained from the Gibbs energy, $G$,
\begin{equation}
G(T,p)=\int_{V_m}^{V{_{m,u}}}p(V_m^{'},T,\chi_i)\mathrm{d}V_m^{'}+pV_m-RT+
\sum_i \chi_i\left[G_\alpha^\mathrm{ref}+RT\mathrm{ln}\left(\chi_i\right)\right],
\end{equation}
where the superscript ``$\mathrm{ref}$'' indicates the {\it low-pressure} 
reference condition~\cite{prausnitz} corresponding to $p^\mathrm{ref}=100\,\mathrm{kPa}$, and
 $V_{m,u}=RT/(p^\mathrm{ref})$:
\begin{equation}
h=G-T\left.\left(\dfrac{\partial G}{\partial T}\right)\right|_{p,\chi}
=h^\mathrm{ref}+pV_m-RT+K_1\left(a-T\ppder{a}{T}\right),
\label{eq:prh}
\end{equation}
and
\begin{equation}
c_p=\left.\left(\ppder{h}{T}\right)\right|_{p,\chi}=
c_p^\mathrm{ref}-T\dfrac{\left(\partial p/ \partial T\right)_{V_m,\chi}^2}{\left(\partial p/ \partial V_m\right)_T}
-R-T\dfrac{\partial^2 a}{\partial T^2}K_1,
\label{eq:prcp}
\end{equation}
\begin{equation}
K_1=\frds{1}{2\sqrt{2}b}\mathrm{ln}\left[\frds{V_m+\left(1-\sqrt{2}\right)b}{V_m+\left(1+\sqrt{2}\right)b}\right].
\end{equation}
In equations~(\ref{eq:prh}) and~(\ref{eq:prcp}) the partial derivatives of $a$ with respect 
to the temperature are evaluated analytically, see Ref~(\citen{prausnitz}) for their expressions.

\subsection{Transport properties}
\label{par:transp-model}
Two advanced models have been
implemented in this work to evaluate correctly the viscosity~\cite{vischung} and the thermal conductivity~\cite{elyhanley83},
which strongly depend on the pressure at the thermodynamic conditions of interest here.

Cho and Chung~\cite{vischung}, starting from a base formulation valid at 
low pressures, developed an expression for the viscosity which is valid at high 
(supercritical) pressure and low (transcritical) temperature:
\begin{equation}
\mu = \mu^* \, \dfrac{36.344 \, \sqrt{M_w \, T_c}}{V_{m,c}^{2/3}},
\end{equation}
see Ref~(\citen{reidbook}) for the meaning of the symbols and for further details.

Ely and Hanley~\cite{elyhanley83} used Eucken's splitting of the 
thermal conductivity into contributions from the interchanges of both translational and internal energies~\cite{reidbook},
together with a corresponding state method using methane as the 
reference component to estimate the thermal conductivity of non-polar fluids over a wide range of 
densities and temperatures:
\begin{equation}
\lambda = \lambda_m^{**} + \sum_i \, \sum_j \, \chi_i \, \chi_j \, \lambda_{ij} \, ,
\end{equation}
see again Ref~(\citen{reidbook}) for the meaning of the symbols and for further details.

\subsection{Flow equations and numerical solution procedure}

The numerical method developed in Ref~(\citen{CutroneCAF09}) has been employed to solve the steady-state RANS equations with $k$-$\omega$ turbulence closure. 
For an axisymmetric multi-component reacting compressible flow the system of the governing equations can be written as:
\begin{equation}
\label{floweq}
{\partial_t \vec Q+\partial_{x} (\vec E-\vec E_{\nu})+\partial_{y} (\vec F-\vec F_{\nu})= \vec S},
\end{equation} 
where $t$ is the time variable; $x$ and $y$ are the axial and the radial coordinate, respectively;
$k$ and $\omega$ are the turbulence kinetic energy and its specific dissipation rate;  
$\vec Q$=($\overline \rho$,$\,\overline \rho \widetilde u_x$,$\,\overline \rho \widetilde u_y$,$\,\overline \rho \widetilde H-p_t$,$\,\overline \rho  k$, $\,\overline \rho \omega$,$\,\overline \rho \widetilde R_n$) is the 
vector of the conserved variables; $\overline \rho$, $(\widetilde u_x,\widetilde u_y)$, $\widetilde H$ indicate the Reynolds-averaged value of density, the Favre-averaged values of velocity components and specific total enthalpy given by $\widetilde H =\widetilde h+\frac{1}{2}(\widetilde{u}_x^2+\widetilde{u}_y^2)+\frac{5}{3}k$ with $\widetilde h$ accounting for the species enthalpy per unit mass, respectively; 
$\widetilde R_n$ is a generic set of conserved variables related to the combustion model.
In this framework, $\widetilde R_n$ is the set of independent variables of the flamelet model, namely, $\widetilde Z$, ${\widetilde{Z''^2}}$,
$\widetilde C$, ${\widetilde{C''^2}}$ (see the following subsection); 
$\vec E$, $\vec F$, and  $\vec E_v$, $\vec F_v$ are the inviscid and viscous flux vectors~\cite{dsthesis}, respectively; $\vec S$ is the vector of the source terms.\\
The heat flux in the total energy equation, namely $q=-\rho c_p D_T\nabla T+\sum_{n=1}^{N_s} \rho V_nY_nh_n$, is composed by two terms since the Dufour effect is neglected, $D_T$ is the thermal diffusivity and $c_p$ the specific heat at constant pressure. The mass diffusion term is treated with the Fick's law considering $V_n=-D_{n,mix}\frac{\nabla Y_n}{Y_n}$, assigning a mixture diffusivity, $D_{n,mix}$, to each species. 
A cell-centered finite volume space discretization is used on a multi-block structured mesh. The convective and viscous terms are discretized by the third-order-accurate Steger and Warming~\cite{steger} flux-vector-splitting scheme and by second-order-accurate central differences, respectively. An implicit time marching procedure is used with a factorization based on the diagonalization procedure of Pulliamm and Chaussee~\cite{pulliam}, employing a scalar alternating direction implicit (ADI) solution procedure~\cite{buelow97}. Steady flows are considered and the ADI scheme is iterated in the pseudo-time until a residual drop of at least five orders of magnitude for all of the conservation-law equations~\eqref{floweq} is achieved. Characteristic boundary conditions for the flow variables are imposed at inflow and outflow points, whereas no slip and adiabatic conditions are imposed at walls; $k$, $\omega$, and $\tilde R_n$ 
are assigned at inflow points, whereas they are linearly extrapolated at outflow points.
At solid walls, $k$ is set to zero and $\omega$ is evaluated as proposed by Menter and Rumsey~\cite{menter}:
\begin{equation}
\label{eq:menter}
\omega = 60 \, \dfrac{\nu}{0.09 \, y_{n,1}^2 },
\end{equation}
where $y_{n,1}$ is the distance of the first cell center from the wall; the homogeneous Neumann
boundary condition is used for $\tilde R_n$ (non-catalytic wall).
Finally, symmetry conditions are imposed at the axis.

\subsection{Turbulent FPV transport equations}

For the case of turbulent flames, the solution of the SLFE, namely equation~\eqref{phi}, is expressed in terms of the Favre averages of $Z$ and $C$ and of their variances. Using Model~A, 
one can tabulate all chemical quantities in terms of $\widetilde Z$, $\widetilde{Z^{''2}}$ and $\widetilde C$, since the model is independent of $\widetilde{C^{''2}}$. On the other hand, Model~B expresses $\phi$ also in terms of $\widetilde{C^{''2}}$ 
and therefore an additional transport equation needs to be solved. 
In this case, the transport equations for the combustion model (included in equation~(\ref{floweq})) are written as:
\begin{eqnarray}
\label{zmean}
 \partial_t(\overline{\rho}\widetilde{Z})+\vec\nabla\cdot(\overline{\rho}\widetilde{\vec u}\widetilde{Z})&=&
\vec\nabla\cdot\Bigl[\bigl( D+
D_{\widetilde{Z}}^t\bigr)\overline{\rho}
\vec\nabla\widetilde{Z}\Bigr],\\
\label{zvar}
\partial_t(\overline{\rho}\widetilde{Z''^2})+\vec\nabla\cdot(\overline{\rho}\widetilde{\vec u}\widetilde{Z''^2})&=&
\vec\nabla\cdot\Bigl[\bigl( D+D_{\widetilde{Z''^2}}^t\bigr)\overline{\rho}\vec\nabla\widetilde{Z''^2}\Bigr]-\overline{\rho}\widetilde{\chi}+2\overline{\rho}D_{\widetilde Z}^t(\vec\nabla\widetilde{Z})^2,\\
\label{cmean}
\partial_t(\overline{\rho}\widetilde{C})+\vec\nabla\cdot(\overline{\rho}\widetilde{\vec u}\widetilde{C})&=&
\vec\nabla\cdot\Bigl[\bigl( D+D_{\widetilde{C}}^t\bigr)\overline{\rho}\vec\nabla\widetilde{C}\Bigr]+\overline{\rho}\overline{\dot\omega_C},\\
\label{cvar}
\partial_t(\overline{\rho}\widetilde{C''^2})+\vec\nabla\cdot(\overline{\rho}\widetilde{\vec u}\widetilde{C''^2})&=&
\vec\nabla\cdot\Bigl[\bigl( D+D_{\widetilde{C''^2}}^t\bigr)\overline{\rho}\vec\nabla\widetilde{C''^2}\Bigr]-\overline{\rho}\widetilde{\chi}_C+2\overline{\rho}D_{\widetilde C}^t(\vec\nabla\widetilde{C})^2+2\overline{\rho}\widetilde{C''\dot\omega''_C},
\end{eqnarray}
where $\widetilde{\chi_C}$ is modeled in terms of $\widetilde{Z''^2}$ and $\widetilde{C''^2}$~\cite{IhmeCF2}, namely $\widetilde{\chi_C}=\frac{\widetilde{Z''^2}\chi}{\widetilde{C''^2}}$, $D$ is the diffusion coefficient for all of the species, given as $D=\nu/Pr$ assuming a unity Lewis number; $\nu$ and $Pr$ are the kinematic viscosity and the Prandtl number, respectively; $D_{\widetilde Z}^t=D_{\widetilde{Z^{''2}}}^t=D_{\widetilde C}^t=D_{\widetilde{C''^2}}^t=\nu_t/Sc_t$ are the turbulent mass diffusion coefficients, ${Sc}_t$ being the turbulent Schmidt number; finally, $\dot{\omega}_{C}$ is the source term for the progress variable precomputed and tabulated in the flamelet library. The gradient transport assumption for turbulent fluxes is used and the mean scalar dissipation rate, $\widetilde{\chi}$ and $\widetilde{\chi}_C$, appear as a sink term in equations~\eqref{zvar} and \eqref{cvar}, respectively.\\
At every iteration, the values of the flamelet variables are updated using equations~\eqref{zmean}-\eqref{cvar} and the Favre-averaged thermo-chemical quantities are computed, using equation \eqref{media}. Such solutions provide the mean mass fractions which are used to evaluate all of the transport properties of the fluid, namely the molecular viscosity, the thermal conductivity and the species diffusion coefficients.\\

\section{MASCOTTE V03: H$_2$/LO$_2$ supercritical combustion}
The MASCOTTE cryogenic combustion test facility was developed by ONERA
to study fundamental processes in cryogenic combustion involving
propellants such as liquid oxygen and gaseous hydrogen. In particular, the test-case RCM-3 2001~\cite{rcm013},
dealing with the super-critical H$_2$/LO$_2$ combustion problem,
has been chosen here as a suitable test for investigating: 1) the influence of the real-gas model 
on the simulated combustion phenomenon;
2) the importance of using an accurate chemical kinetic scheme within the flamelet model;
3) the influence of the presumed probability function model on the flame structure.

The injector consists of an inner diverging duct for the oxygen with inlet diameter of 3.6 mm  and 
outlet diameter of 5 mm. Hydrogen is injected coaxially,
through an annular duct with inner and outer diameters equal to 5.6 mm and 10 mm, respectively.
The injector length is equal to 50 mm so that a fully developed turbulent flow
is obtained at the injector exit. The combustion chamber has a $50\times 50$~mm$^2$ square section, the edge length being 50 mm.
In order to perform an axisymmetric simulation,
the combustion chamber is modeled as a cylinder 
with a radius of 28.21 mm such as to preserve the chamber section area. 

The chamber pressure was held at 
6 MPa, higher than the critical pressure for both oxygen and hydrogen 
(5.043 MPa and 1.313 MPa, respectively). Liquid oxygen is injected at the temperature of 85 K, corresponding to a density
of 1177.8 kg/m\textsuperscript{3}, 
whereas hydrogen is injected at a temperature of 287~K. 
\begin{figure}[h!]
\centering
\includegraphics[bb= 90 78 721 422, clip,width=0.8\textwidth]{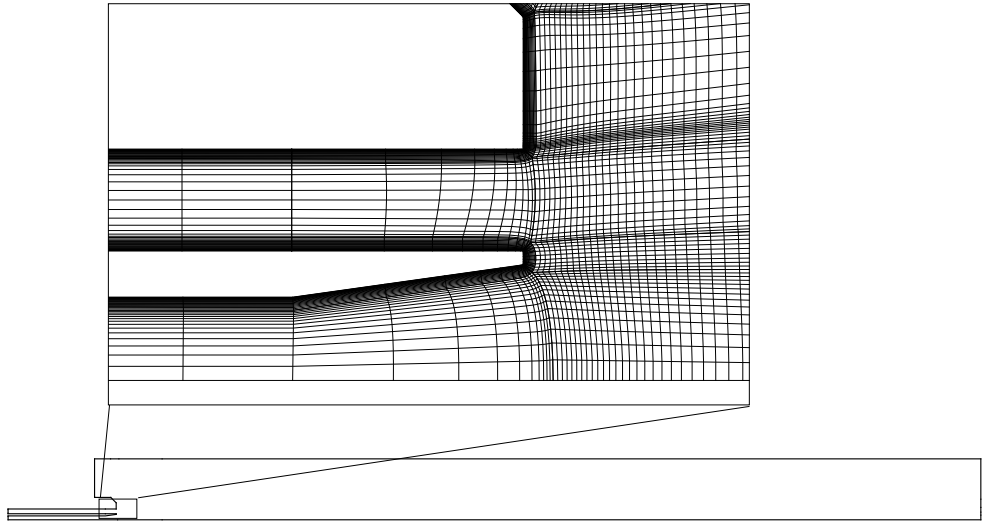}
\caption{Reference geometry for Test RCM-3 2001 and a detail
of the computational grid for the injectors.} 
\label{fig:rcm01-3grid}
\end{figure}
A computational grid with about 18,000 cells distributed among 24 structured blocks has been used. A local view
of the grid is given in figure \ref{fig:rcm01-3grid}.
Table~\ref{tab:rcm01-3} provides the inlet boundary conditions.
\begin{table}[!tb]
\centering
\caption{Conditions for RCM-3 2001 test-case.}
\label{tab:rcm01-3}
\begin{tabular}{l c c c c c c}
\hline
& $\dot{m}$ (kg/s) & $T$ (K) &$\rho$ (kg/m\textsuperscript{3})& $u$ (m/s) &Tu$_{in}$ (-)& $\lambda_{T,in}$, mm\\
LO$_2$ &0.1& 85& 1177.8 & 4.35& 5\% &4\\
H\textsubscript{2} & 0.07 & 287 & 5.51 & 236 & 5\% &4\\
\hline
\end{tabular}
\end{table}

For this test case, a limited experimental dataset for comparison is available.
Quantitative experimental profiles of temperature~\cite{habib2006} are available at three axial locations ($x=$0.015, 0.05 and 0.1m), with a 
quite poor resolution (four points at $r=$ 0.004, 0.008, 0.012 and 0.016m).
On the other hand, any quantitative experimental data are available for chemical species.
Additionally, a LIF image data~\cite{mascotteIAC00} of the hydroxyl radical is used for qualitative
comparison.

Steady flamelet calculations have been performed 
with the chemical kinetic scheme provided by Li \emph{et al.}~\cite{Lietal2004}, from now on
called Li-scheme, who developed
a scheme consisting of 25 elementary reactions and 9 species, which was proven accurate for operating
pressures up to 9~MPa.
In the present study,
the chemistry library is
discretized with 125 uniformly distributed points in the $\wt Z$ and $\wt C$
directions, and 25 points in the $\wt \zvar$ and $\wt {C''^2}$ direction.

\subsection{Gas model effects}

\label{sec:FSHP}
The system of the flamelet equations is
solved here employing either the ideal-gas EoS, with state-of-the-art models
for the temperature dependence of the thermodynamic and transport properties~\cite{cutronePhD}, or the real-gas EoS
with the advanced transport properties described in sections~\ref{par:thermo-model} and \ref{par:transp-model}.

\begin{figure}[!t]
\centering
\includegraphics[bb=71 2 577 744,clip,angle=-90,width=0.45\textwidth]{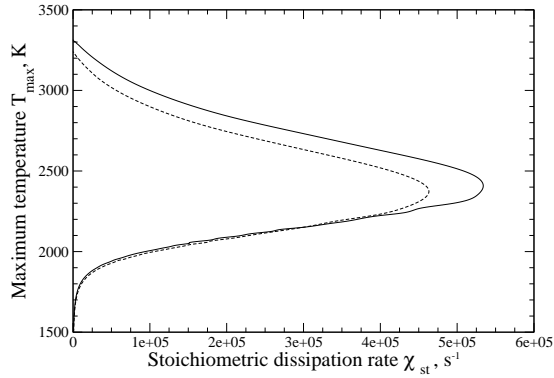}
\caption{
S-shaped curve obtained with ideal- (solid line) and real-gas (dashed line) models.}
\label{fig:scurve_gasmod}
\end{figure}

\begin{figure}[!t]
\centering
\subfigure[\label{subfig:analisi-prod}]{\includegraphics[bb=51 17 460 670, clip,angle=-90,width=0.45\textwidth]{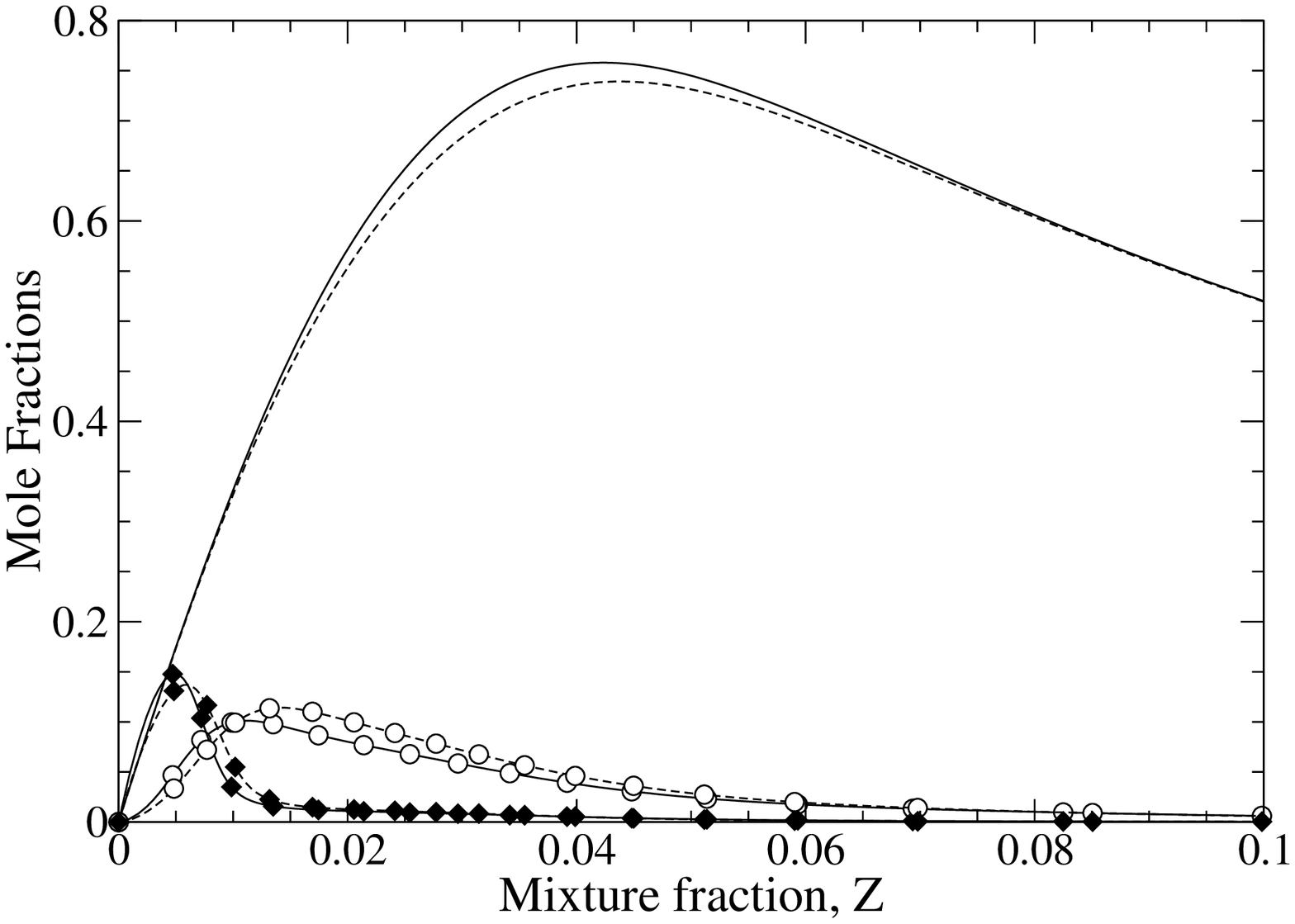}}
\subfigure[\label{subfig:analisi-rad}] {\includegraphics[bb=51 17 460 670, clip,angle=-90,width=0.45\textwidth]{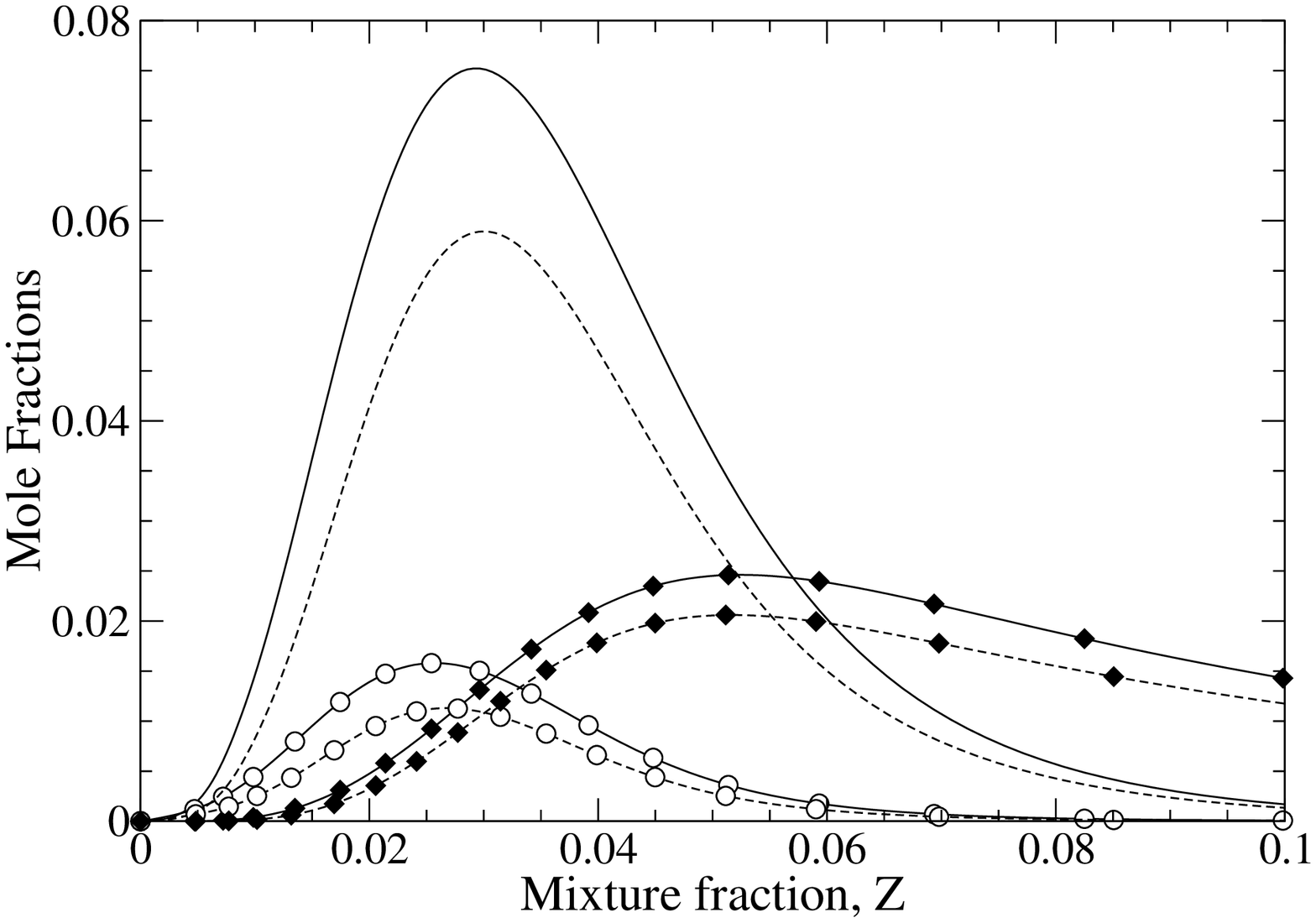}}
\caption{
Ideal-gas (solid line)  and real-gas (dashed line) solutions at 
\mbox{$\chi_\mathrm{st}=10^5$ s$^{-1}$.}
\subref{subfig:analisi-prod} final products:
$\mathrm{H_2O}$ (plain line), $\mathrm{HO_2}\cdot 100$ (open circles), $\mathrm{H_2O_2}\cdot 100$ (solid diamonds);
\subref{subfig:analisi-rad} radical species:
$\mathrm{OH}$ (plain line), $\mathrm{O}$ (open circles), $\mathrm{H}$ (solid diamonds).
}
\label{fig:analisi_species}
\end{figure}

The influence of the real-gas model on the flame structure, and consequently on the flamelet calculation, is quite remarkable,
as just pointed out by some authors~\cite{CutroneCAF09,kimkimkim11}.
The application of the real-gas model reduces the maximum 
temperature, due to a lower heat release, and consequently predicts
lower peaks for almost all of the species concentrations.
For an $\H2$/L$\o2$ combustion at a 6 MPa pressure,
the two S-shaped curves computed by using 
either the ideal- or the real-gas models are provided in figure~\ref{fig:scurve_gasmod}.
\begin{figure}[!t]
\centering
\subfigure[\label{subfig:analisi-temp}]{\includegraphics[bb=51 17 460 670, clip,angle=-90,width=0.45\textwidth]{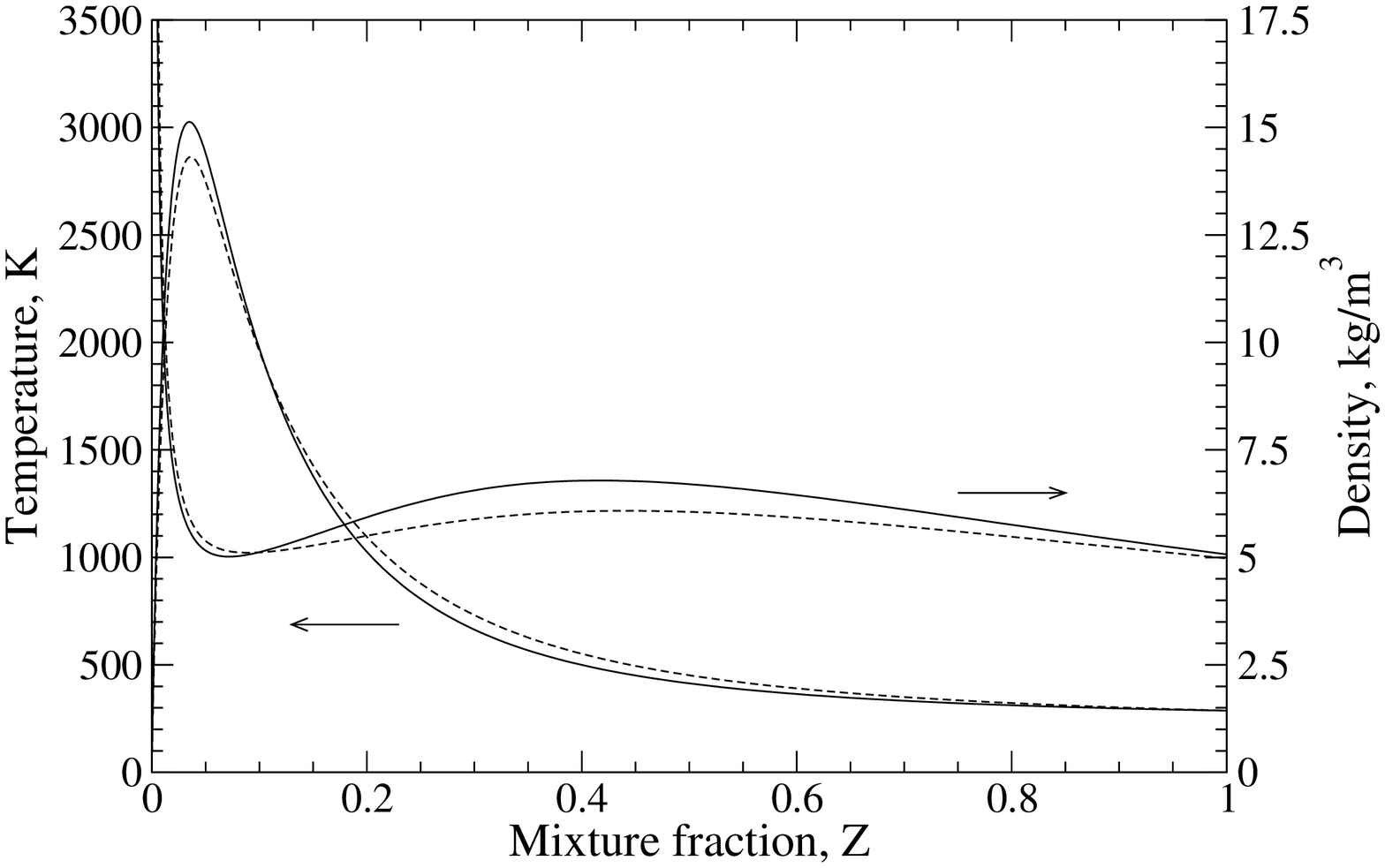}}
\subfigure[\label{subfig:analisi-hr}]  {\includegraphics[bb=51 17 460 670, clip,angle=-90,width=0.45\textwidth]{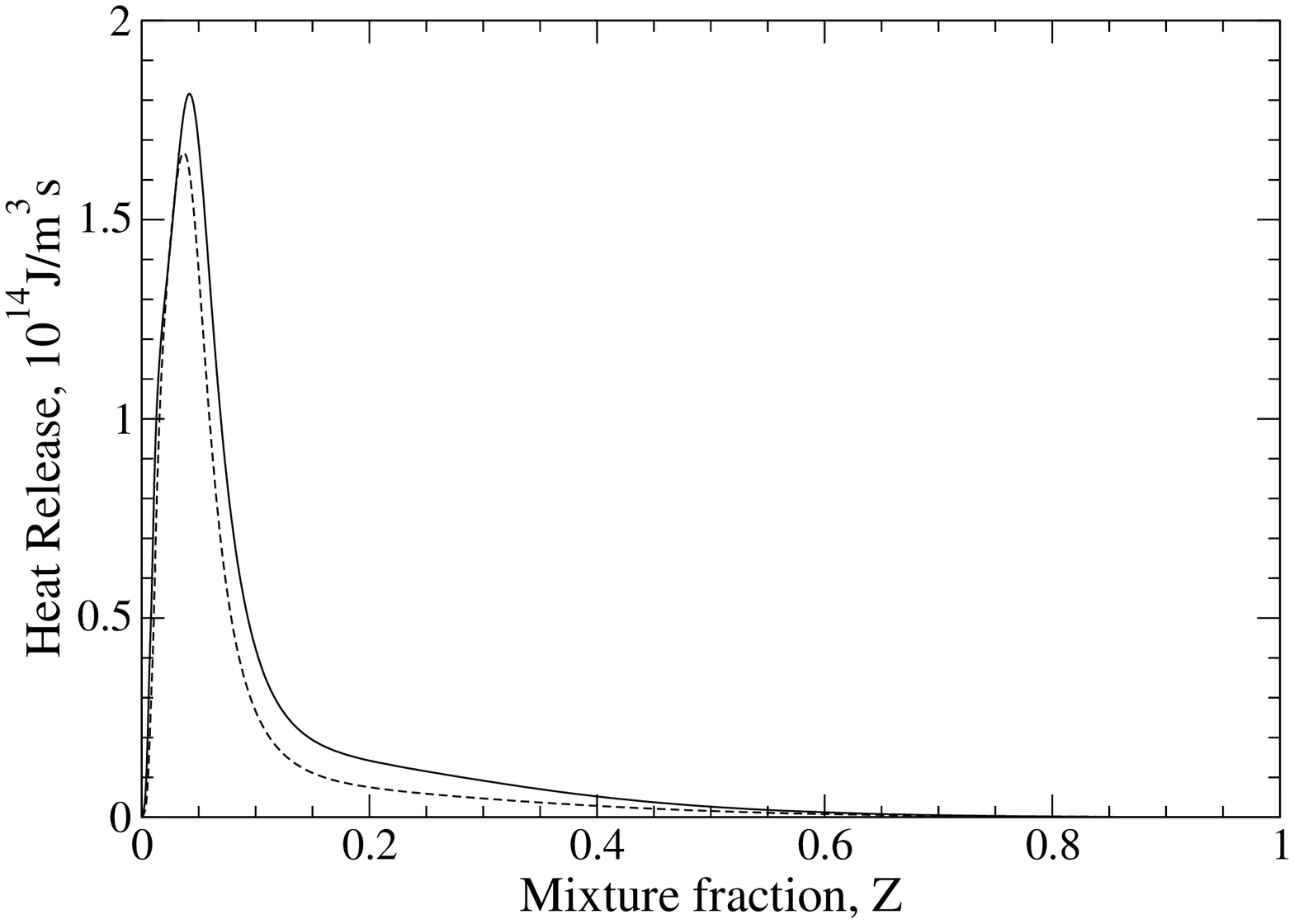}}
\caption{
Ideal- (solid lines)  and real-gas (dashed lines) solutions at
\mbox{$\chi_\mathrm{st}=10^5$ s$^{-1}$.}
\subref{subfig:analisi-temp} temperature and density;
\subref{subfig:analisi-hr} heat release.}
\label{fig:analisi_temp_hr}
\end{figure}
Due to the higher dissipation rate, the ideal-gas model provides a 
higher hydrogen consumption rate along the entire stable upper branch, resulting in higher flame temperatures,
as shown in figure~\ref{fig:analisi_temp_hr}\subref{subfig:analisi-temp}. 
Accordingly, a higher heat release is predicted by the  ideal-gas flamelet with respect to the
real-gas one, with the peak shifted towards the oxidizer side, see 
figure~\ref{fig:analisi_temp_hr}\subref{subfig:analisi-hr}.
Therefore, the lower fuel consumption rate for the real-gas flamelet leads to lower species concentrations,
as shown in figures~\ref{fig:analisi_species}\subref{subfig:analisi-prod}-\subref{subfig:analisi-rad}:
the mole fractions for the final combustion products and radical species are lower for the real-gas 
flamelet for all species except $\mathrm{HO}_2$.

These sets of flamelet have been integrated accordingly to both Model~A and Model~B approaches, but only the first
will be used in this analysis on real-gas effects.

Firstly, a simulation has been performed using the ideal-gas 
model for both the equation of state and the flamelet model. The ideal-gas
equation of state fails in the prediction of the density of
oxygen at the injection temperature of 85 K and a chamber pressure of 6 MPa:
it provides a value of 271.7
kg/m\textsuperscript{3} versus an experimental value~\cite{webbook} of 1177.8 kg/m\textsuperscript{3}.
The main issue of such model is that the computed oxygen mass-flow rate drops to 0.023~kg/s.

\begin{figure}
\centering
\textbf{Effects of real-gas model}
\centering
\includegraphics[bb=-90 196 800 570,clip,width=\textwidth]{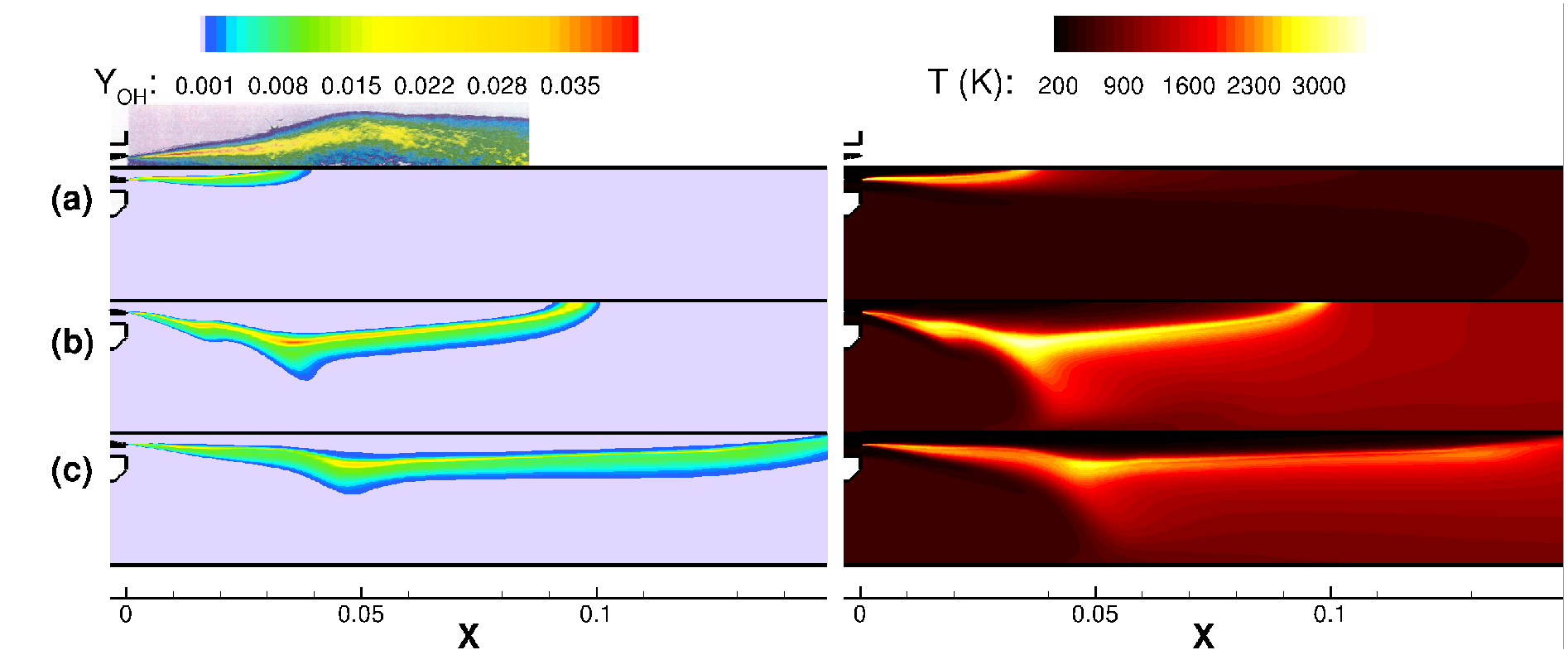}
\caption{OH mass fractions (left) and temperature (right): (top-left) Abel-transformed-emission image, 
(a) Ideal-gas EoS -- Ideal-gas flamelet model, (b) Real-gas EoS -- Ideal-gas flamelet model, (c) 
Real-gas EoS -- Real-gas flamelet model.}
\label{fig:realvsideal_c}
\end{figure}

\begin{figure}
\centering
\textbf{Effects of real-gas model}
\centering
\includegraphics[angle=-90,width=\textwidth]{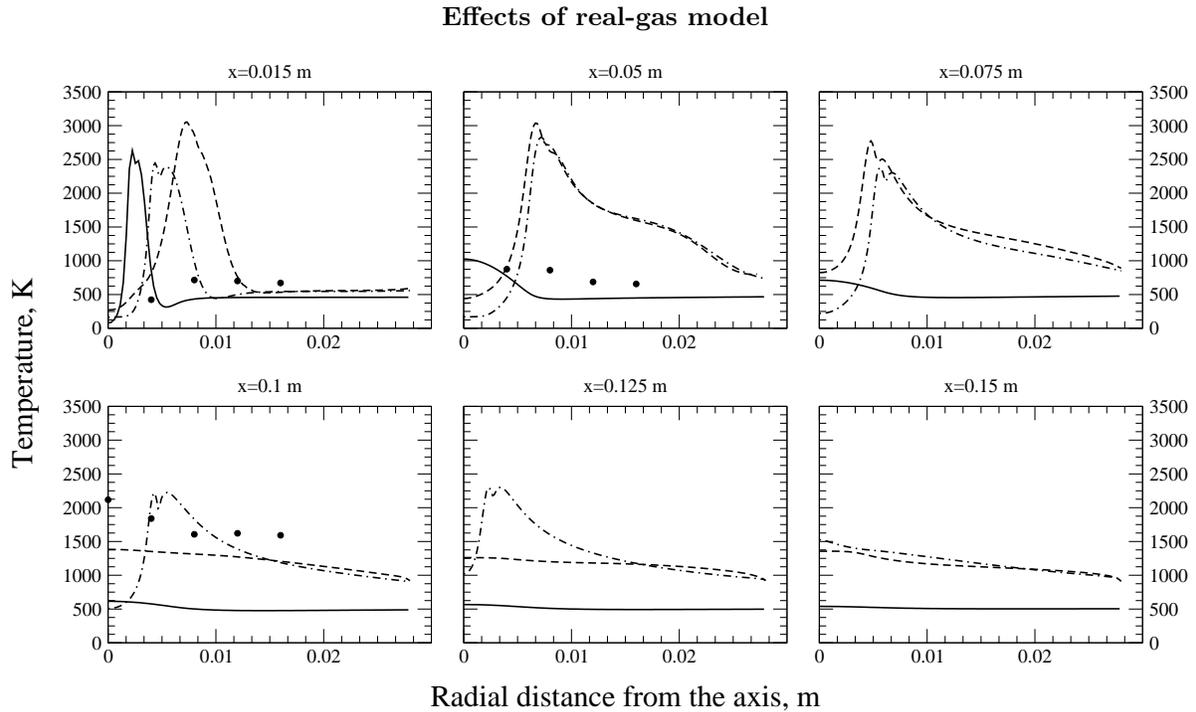}
\caption{Radial temperature distributions at several axial locations:   Ideal-gas EoS -- Ideal-gas flamelet (solid line),
Real-gas EoS -- Ideal-gas flamelet (dashed line), Real-gas EoS -- Real-gas flamelet (dashed-dotted line) models }
\label{fig:realvsideal_p}
\end{figure}

Figure~\ref{fig:realvsideal_c} provides a comparison between numerical results and experimental data
for OH mass fractions (left), and a comparison between numerical results for temperature (right).
As expected, the ideal-gas model (a) does not appear
adequate for the present test-case, providing a very short flame, essentially due to the
reduced oxygen mass-flow-rate, and consequently a different oxygen/fuel ratio.

Then, the influence of the gas-model
onto the derivation of the flamelet library has been investigated.
The influence of the gas-model on the temperature field is quite remarkable,
as shown in figure~\ref{fig:realvsideal_c}, which provides the results
obtained using the ideal- and real-gas models in the flamelet approach, indicated by label (b) and (c), respectively.
It appears that, due to the higher dissipation rate, the ideal-gas flamelet model
provides a higher hydrogen consumption rate and, therefore, a shorter flame, which seems closer to the experimental result
that provides a closed flame at about $x=$ 0.09m. However,  the ideal-gas flamelet model exhibits a higher spreading angle,
in particular close to the injector, where the flame should be a thinner layer confined around the oxygen core.

A more detailed comparison is given in figure~\ref{fig:realvsideal_p}, showing the radial
temperature distributions at six positions along the axis.
Close to the injector ($x=$ 0.015m), the real-gas flamelet model results are in reasonable 
agreement with the experimental data except the hot flame zone close to the axis, where CARS data are not 
sufficiently resolved to capture the flame layer. On the other hand, the ideal-gas flamelet model 
provides a wider flame due to the higher spreading angle observed.
Slightly apart from the injector ($x=$ 0.05m) both the ideal- and real-gas flamelet model overestimate the temperature in the 
hydrogen-rich flame zone ($r=$ 0.012 and 0.016m), but again, any indication on the position of the flame
front (somewhere between $r=$ 0.004 and 0.012m) is provided by experimental data which exhibits a flat profile of about 1000K.
Finally, radial distributions confirm that the ideal-gas flamelet model provides a
faster combustion, leading to low values of the temperature gradient, characteristic of the
post-combustion diffusion process, already at $x=0.1~m$, where experimental profile indicates that the hot flame zone could 
be placed in the proximity of the jet core.

\subsection{Kinetic scheme effects}
Here, in order to assess the role of the basic chemical kinetic scheme employed for the definition
of the steady flamelet equations, three alternative schemes have been considered.
Two of them are simplified models proposed by Jachimowski~\cite{jachi} and Marinov\cite{marinov95}, from now on
called Jachimowski- and Marinov-scheme, respectively; the former is a reduced scheme with 7 reactions and 7 species, whereas the latter is 
a global scheme (3 species).
The third one is an additional detailed schemes provided by Warnatz~\cite{warnatz88} (from now on called Warnatz-scheme).

The performance of these mechanisms are at first evaluated in a 
shock tube laminar test for which experimental data, namely ignition delay times,
are available~\cite{petersen95}.
The AURORA code\cite{aurora}, was used to simulate experimental
conditions in the shock tube. The flow is assumed
zero-dimensional, laminar and well mixed.
The operating pressure is $p=6.5$ MPa and the initial composition is set to:
$Y_{\o2}=0.01$, $Y_{\H2}=0.02$, and $Y_{\mathrm{Ar}}=0.97$, whereas
the initial temperature ranges from 1000 to 2000 K.
The ignition delay time is defined as the time when
$\d [\oh]/\d t$ reaches its maximum, where $[\oh]=\rho Y_\oh/\mathcal{M}_\oh$ indicates the 
molar concentration of the species $\oh$. As a consequence of this definition,
the ignition delay time cannot be calculated with the Marinov-scheme, being the OH species concentration not calculated.
\begin{figure}[pt]%
\centering%
\includegraphics[angle=-90,bb=71 17 592 744,clip,width=0.45\textwidth]{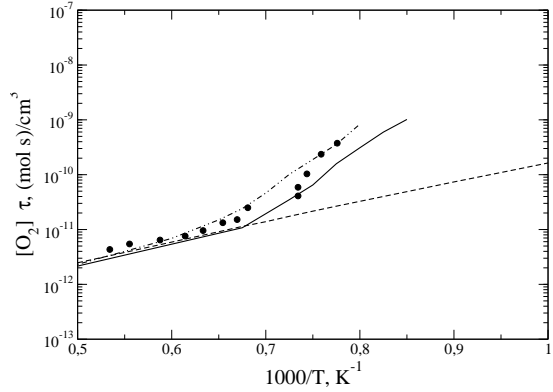}%
\caption{Ignition delay of $\H2/\o2/\mathrm{Ar}$ mixtures in shock tubes. Li-scheme (dashed-dotted-dotted line), 
Warnatz-scheme (solid line), and Jachimowski-scheme (dashed line).}%
\label{fig:igndel}%
\end{figure}%
In  figure~\ref{fig:igndel} the ignition delay times of $\H2/\o2/\mathrm{Ar}$ mixtures are reported
for the Li-, Warnatz- and Jachimowski-scheme. It appears that
(1) the reduced Jachimowski-scheme provides very short ignition delay at low initial 
temperatures, diverging from the experimental at  about 1400 K;
(2) the detailed schemes are able to capture the rapid increase of the
ignition delay at lower temperature, that as indicated by Refs~(\citen{petersen95,skinner65}),
is due to the increased chain-termination mechanisms.

Then, the considered schemes have been used for the generation of flamelet libraries and
2D axisymmetric CFD computation have been performed.
All the results have been obtained using the real-gas flamelet model and the turbulent combustion Model~A.

\begin{figure}
\centering
\textbf{Kinetic scheme effects}
\centering
\includegraphics[width=\textwidth]{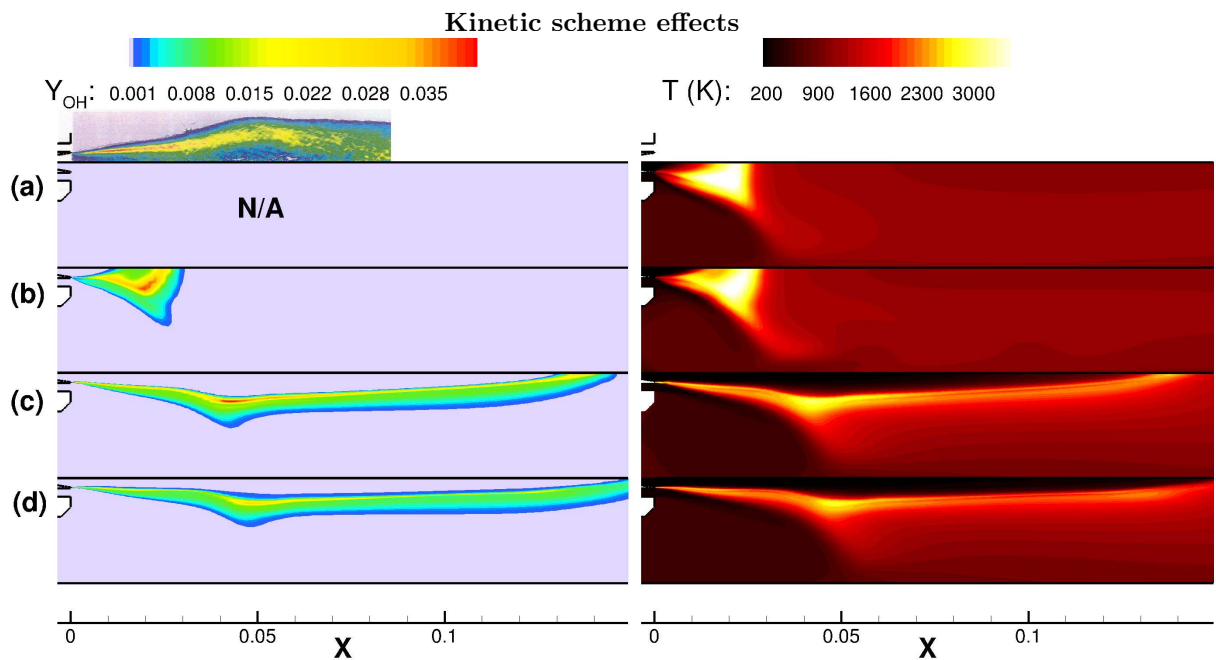}
\caption{OH mass fractions (left) and temperature (right): (top-left) Abel-transformed-emission image,
(a) Marinov-scheme, (b) Jachimowski-scheme
(c) Warnatz-scheme, (d) Li-scheme.}
\label{fig:redvsfull_c}
\end{figure}
\begin{figure}
\centering
\textbf{Kinetic scheme effects}
\centering
\includegraphics[angle=-90,width=\textwidth]{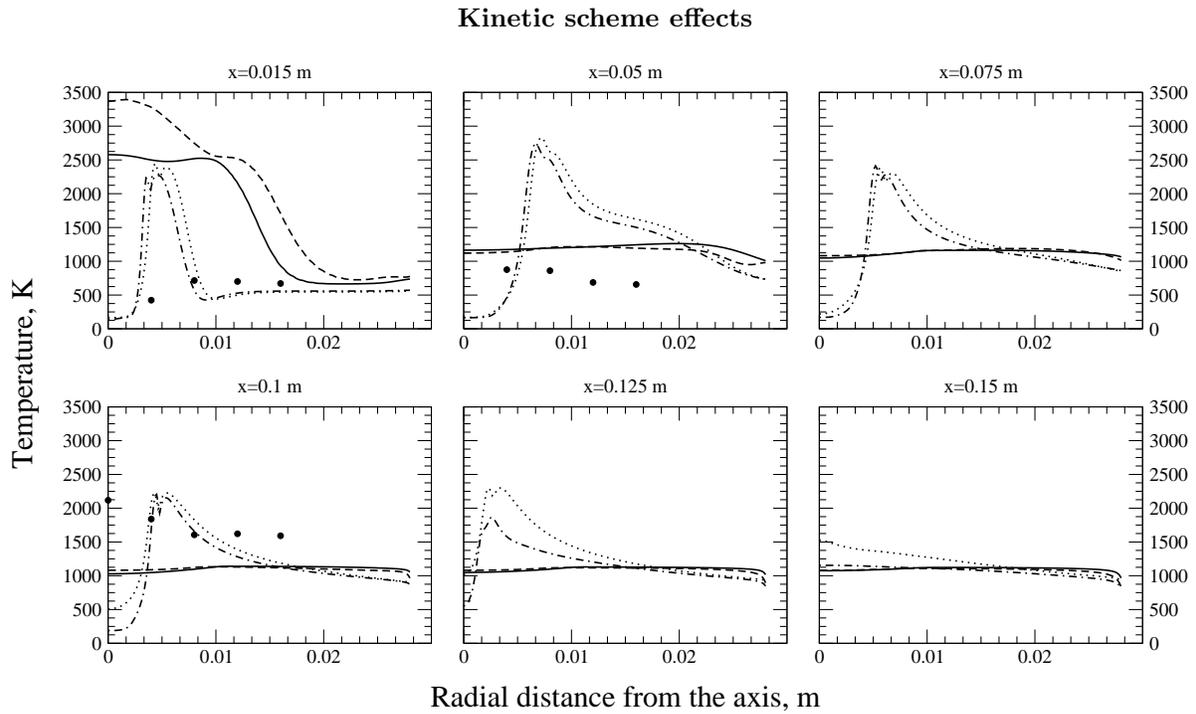}
\caption{Radial temperature distributions at several axial locations:   Marinov-scheme (solid line),
Jachimowski-scheme (dashed line), Warnatz-scheme (dash-dotted line), Li-scheme (dotted line) models.}
\label{fig:redvsfull_p}
\end{figure}

Figure~\ref{fig:redvsfull_c} shows OH mass fractions and temperature
contours obtained using the Marinov-, Jachimowski-,  and Warnatz-schemes, respectively, the results
computed using the Li-scheme being given for comparison, too. The results obtained with the simplified scheme are unacceptable,
whereas the two detailed schemes qualitatively reproduce the experimental distributions.
A more detailed comparison is provided in figure~\ref{fig:redvsfull_p} which shows the temperature
distributions at several axial locations obtained using the four kinetic schemes.
The results obtained using the two simplified schemes indicate again
that the flame is very short with respect to the experimental data as well as to the numerical results obtained
using either detailed scheme, which appear to be more accurate.

In fact, both simplified schemes predict a barely appreciable ignition delay, the combustion
developing as soon as the reactants come into contact.
Therefore, the combustion takes place very close to the injector and the reactants
cannot be transported by the flow further downstream. Such results confirm that, especially for
high-pressure combustion processes, a detailed kinetic scheme is warranted.
It is noteworthy that different temperature
distributions are provided by the Li- and  Warnatz-scheme, the latter predicting
a slightly shorter flame with a smaller spreading angle and a thinner reaction zone.

\subsection{Turbulent combustion model effects}
The results shown in the previous sections have been obtained using the flamelet Model~A,
which based on the original flamelet--progress-variable model developed by Pierce \emph{et al.}~\cite{piercemoin2004}.
Here, in order to asses the role of the presumed probability function on high-pressure combustion, 
the flamelet Model~B has been used to simulate the MASCOTTE test case.
All the results of this section have been obtained using the real-gas flamelet model and the Li-scheme.

\begin{figure}
\centering
\centering
\textbf{Turbulent combustion model effects}
\centering
\includegraphics[width=\textwidth]{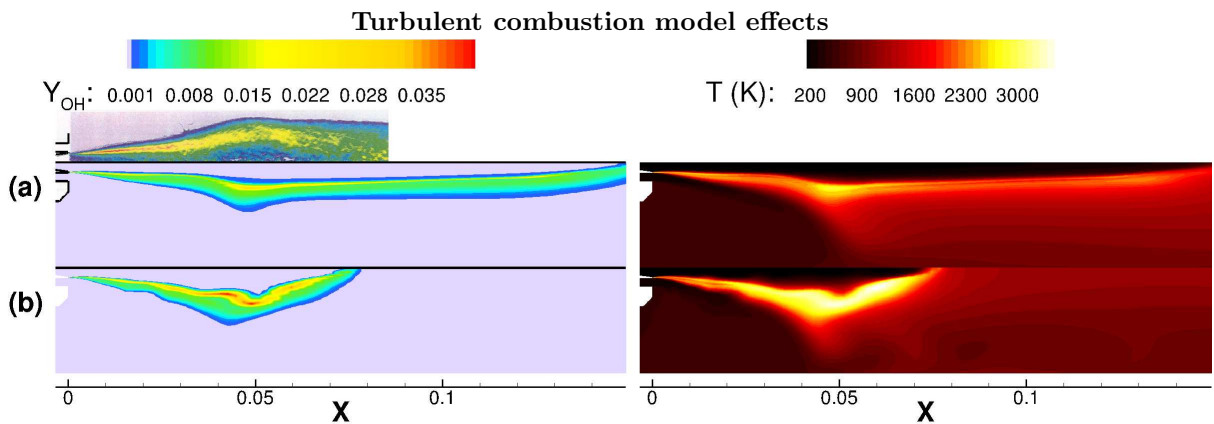}
\caption{OH mass fractions (left) and temperature (right): (top-left) Abel-transformed-emission image,
(a) Model~A,  (b) Model~B.}
\label{fig:smldvsfpv_c}
\end{figure}
\begin{figure}
\centering
\textbf{Turbulent combustion model effects}
\centering
\includegraphics[angle=-90,width=\textwidth]{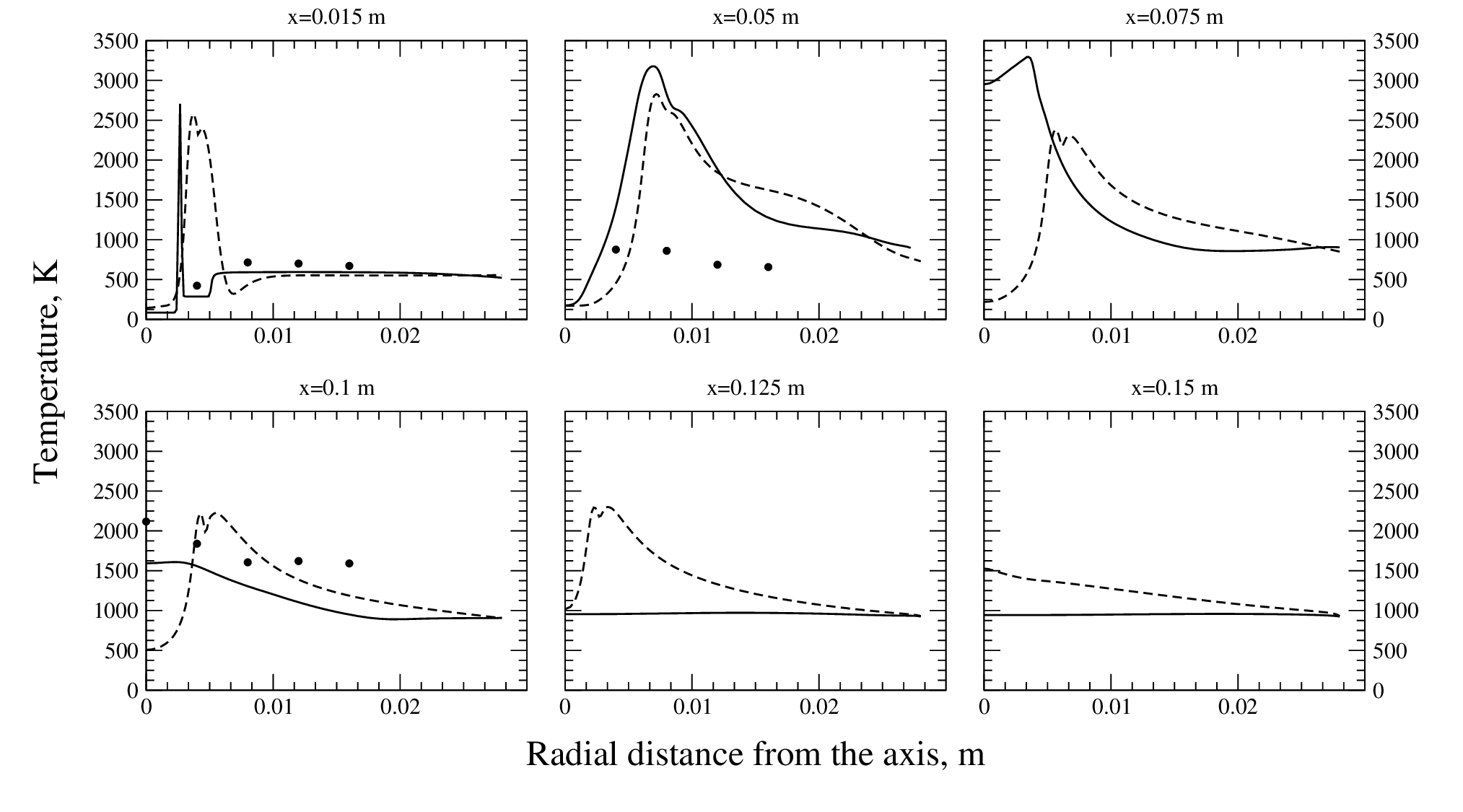}
\caption{Radial temperature distributions at several axial locations:   Model~A (dashed line),
Model~B (solid line).}
\label{fig:smldvsfpv_p}
\end{figure}

Figure~\ref{fig:smldvsfpv_c} shows OH mass fractions and temperature contours obtained using the Model~A and Model~B.

The results obtained by Model~B seems qualitatively in better agreement with experimental data with respect to Model~A.
At first, Model~B shows a shorter flame length with respect to Model~A and this seems to be in accordance with 
experimental observations. Moreover, the characteristic bump of the flame front due to the entrainment of 
cold fluid from the corner vortex is moved backward, likewise to the experimental observation,
indicating that the injection and mixing processes combined with 
combustion are better captured. Last but not least, the flame thickness 
provided by Model~B grows more slowly in the first  part of the flame 
close to the injector, where the flame should be a very thin hot layer 
around the cold L$\o2$ core, whereas it became larger moving slightly forward.

A more detailed comparison is provided in figure~\ref{fig:redvsfull_c} which shows the temperature distributions. 
The results indicate again that the Model~B flame is very thin close to the injectors ($x=$ 0.015m) 
and computed temperature is in very good agreement with experimental data. 
In the section at $x=$ 0.1m, Model~B shows a better agreement with the 
experimental data, the temperature is underestimate but the behavior is well predicted. 
It seems, in fact, that the combustion process is too fast in the last part of the flame.
In conclusion, seems that the more general framework given by Model~B can actually 
improve the prediction capabilities of the combustion model joined with the real gas equation of state. 
The greatest differences are observed considering the two flame shapes and the initial part, 
near to the injector $x\le$0.03 m, due to the better evaluation of the corner vortex.

\subsection{Conclusions}
This paper provides a numerical method based upon RANS equation for the simulation of high-pressure
conditions. The turbulent combustion coupling has been modeled by implementing  flamelet--progress-variable models,
with the $\h2$/L$\o2$ combustion kinetics provided by four kinetic schemes (both reduced and detailed), and thermodynamics by 
the Peng Robinson real-gas equation of state.
Moreover, a general framework for the evaluation of the most probable joint 
distribution of the mixture fraction and the progress variable has been developed and used to generate a flamelet lookup-table.

The MASCOTTE V03 test case has been computed, involving the supercritical combustion of $\h2$/L$\o2$ and the effects of the modeling 
on the results has been analyzed. In particular, the test:
\begin{itemize}
\item allowed for a detailed assessment of the real-gas effects, on both the main flow equations and flamelet calculations;
\item showed the importance of using a detailed kinetic scheme;
\item proposed a more general framework for the derivation of the density probability function for FPV models that could improve actual predictions.
\end{itemize}

\bibliographystyle{aiaa}
\expandafter\ifx\csname natexlab\endcsname\relax\def\natexlab#1{#1}\fi
\bibliography{biblio.bib}

\end{document}